\documentclass[seceq]{ptptex}

\usepackage[dvips]{graphicx,color} 
\usepackage{amssymb} 
\usepackage{latexsym}

\title{%        %You can use \\ for explicit line-break
Relative-Velocity-Dependent Weber-type Models in Electromagnetism%
}

\author{%       %Use \scshape  for the family name
Santosh \textsc{Devasia}%
}

\inst{%     %Affiliation, neglected when [addenda] or [errata]
U. of Washington, Seattle, WA 98195-2600, USA
}

\abst{%         %this abstract is neglected when [addenda] or [errata]
This article reconsiders the 
relative-velocity-dependent approach to modeling electromagnetism  
that was proposed initially 
by Weber before  data from cathode-ray-tube (CRT) experiments was available.
It is shown that identifying the nonlinear, relative-velocity   
terms using CRT data results in a model, 
which  not only captures standard 
relativistic effects in optics, high-energy particles, and gravitation,  
but also 
explains apparent discrepancies between predicted and measured energy (i)~in  
high-energy-particle absorption  experiments and (ii)~in the 
classical $\beta$-ray spectrum of radium-E.
}

\graphicspath{{converted_graphics/}}
\begin{document}

\maketitle

\section{\label{sec_1}Introduction} 
This article develops a Weber-type,   
relative-velocity-dependent electromagnetism model. 
Relative-velocity dependence of  electromagnetism models  was proposed initially 
by Weber~\cite{Weber_assis}  
before  data from cathode-ray-tube (CRT) experiments was available.
In contrast, this article develops a nonlinear relative-velocity-dependent model that is based on 
data from CRT experiments. 
This article addresses two challenges in the use of such 
relative-velocity-dependent models: (i)~to capture both low-velocity and high velocity effects 
in electromagnetism; and (ii)~to maintain model-invariance between inertial reference frames.   
The first challenge is addressed by using the  nonlinearity of the proposed model to 
capture: (a)~low-velocity effects such as the force between two 
current carrying wires; and (b)~high-velocity effects 
such as the mass increase seen in 
CRT experiments.  
The second challenge, to maintain model-invariance between 
different inertial frames, is addressed by accounting for 
the relative-velocity effects in  Lorentz and Maxwell's equations. 
The resulting model not only captures 
relativistic effects in optics, high-energy particles, and gravitation but also 
explains apparent discrepancies between predicted and measured energy in   
(i)~the absorption of high-energy particles in cloud chambers~\cite{Turin_Crane_II} 
and  
~\nocite{Ellis_Wooster_1927,Madgwick,scott_35,Ho_36}
(ii)~the average energy  determination of the $\beta$-ray spectrum 
using magnetic fields.~\cite{Ellis_Wooster_1927}\tocite{Ho_36} \

Measurements of the energy lost by high-energy electrons  due to absorption
in lead, by Crane and co-workers, tend to be more than $50\%$ of the expected value.~\cite{Turin_Crane_II}  \  
Studies of possible electron scattering and potential increase in path length (through the lead) 
could not explain this  discrepancy.~\cite{Slawsky_Crane_IV} \  
The inability to resolve this discrepancy led 
Richardson and Kurie~\cite{Richardson36} to conclude that the cloud-chamber-absorption 
method is not reliable for energy measurements \lq\lq{even 
when applied by careful investigators.}\rq\rq ~
However, the discrepancy in the measured energy can be accounted for by 
the model presented in this article.

The proposed model also 
explains apparent discrepancies between predicted and measured energy in  
classical $\beta$-ray spectrum experiments. Ellis and Wooster~\cite{Ellis_Wooster_1927} 
found the average disintegration energy of 
Radium~E (Ra~E) to be $0.344MeV$ using temperature measurements while 
reporting that the  average energy found by Madgwick from the $\beta$-ray spectrum  
was $0.395MeV$; data from both approaches are presented by Ellis and Wooster.~\cite{Ellis_Wooster_1927} \  
Moreover, Madgwick's data (including the location of the spectrum's peak) was 
confirmed independently   by Ho and Wang.~\cite{Ho_36} \  
The difference in the average energy, between these two different approaches,  
led to the development of several  
correction factors for potential errors in measurements at the lower end of 
the $\beta$-ray spectrum for reconciling the difference.~\cite{Martin_Townsend39,neary_40} \  
In contrast, the predicted average energy of the $\beta$-ray spectrum  using the proposed relative-velocity approach 
matches the value from  temperature measurements by Ellis and Wooster~\cite{Ellis_Wooster_1927}, without the need for correction factors.

\vspace{0.1in}
The proposed relative-velocity model has the following general form for the Lorentz force $F_E$
on a particle of charge $q$ due to an electric field $E$ 
%\begin{eqnarray}
\begin{equation}
F_E  ~ = ~  \left[{\cal N}_{\bot}(v_{rel})\right]  q   E_{\bot} ~ 
+ \left[{\cal N}_{\|}(v_{rel})\right]   q E_{\|} 
\label{nonlinear_Lorentz} 
\end{equation}
%\nonumber \end{eqnarray}
where $ E_{\bot}$ and $E_{\|}$ are the components of the 
field perpendicular and parallel to the relative velocity 
$v_{rel}$ between the field and the particle.  
It is noted that ad-hoc choices of the nonlinearities (${\cal N}_{\bot}, {\cal N}_{\|}$ in equation~~\ref{nonlinear_Lorentz}) 
are not acceptable. For example, 
instead of a velocity dependent increase in mass, $m = m_0/\sqrt{1 - \frac{|v_{rel}|^2}{c^2}}$, 
a reduction of the Lorentz force, such as ${\cal N}_{\bot} = \sqrt{1 - \frac{|v_{rel}|^2}{c^2}}$,  
might be considered  
to match the relativistic, velocity-dependent increase in mass in CRT experiments 
(where $c$ is the speed of light and $m_0$ is the rest mass). However, it is shown that 
such a nonlinearity is not consistent with low-speed effects 
such as 
Ampere's law for the force between two current-carrying wires. 

In this article, the form of 
the perpendicular nonlinearity ${\cal N}_{\bot}$ (in equation~~\ref{nonlinear_Lorentz}) 
is identified using: 
(i)~the relativistic mass increase in CRT experiments; and (ii)~conservation of the field's energy density at low speeds. It is shown that the resulting form of the perpendicular 
nonlinearity ${\cal N}_{\bot}$ 
also uniquely identifies: (i)~the  kinetic energy of a particle; and (ii)~the parallel nonlinearity  
${\cal N}_{\|}$. In addition to  matching the relativistic mass increase in CRT experiments (as expected, since the nonlinearity ${\cal N}_{\bot}$ is identified using CRT observations),  the resulting 
nonlinearity expression also matches the low-speed Ampere's law for the force between two current carrying wires. Furthermore,  the resulting energy expression explains the apparent discrepancies in the 
measured energy  in  high-energy electron absorption~\cite{Turin_Crane_II}  and  
Ra~E disintegration.~\cite{Madgwick} \    
Moreover, it is shown that  similar 
relative-velocity-dependent terms result in an expression for the 
precession of planetary orbits that matches the prediction using general relativity.~\cite{Goldstein_80} \

The second difficulty, to find appropriate transformations to 
relate observations in different inertial frames, can be addressed by the proposed 
relative-velocity-based approach where spatial velocity distributions ($V_E, V_B$) are assigned to the electrical $E$ and magnetic fields $B$. 
It is shown that  Maxwell's equations, when adapted to include these relative-velocity distributions, 
are still co-ordinate invariant. 
The effects of the proposed approach on the 
propagation of light and the explanation of optical 
phenomena are considered in this article. 
Most importantly, the 
relative-velocity approach  
models relativistic effects such as (i)~the 
transverse Doppler effect and (ii)~the convection of light 
by moving media (Fresnel drag).  Thus, the article presents a Weber-type 
relative-velocity dependent modeling approach that: (i)~captures relativistic effects in optics, high-energy particles, and gravitation; and 
(ii)~explains apparent discrepancies in experimental energy measurements.    

%\newpage
\section{Relative velocity approach}
In this section, the nonlinearities ${\cal N}_{\bot}, {\cal N}_{\|}$ (in equation~~\ref{nonlinear_Lorentz}) 
are identified. Low speed effects (energy density invariance) and high speed (CRT) effects are considered in the first two subsections to identify the 
perpendicular nonlinearity ${\cal N}_{\bot}$, which is then used in the third subsection to identify 
expressions for the kinetic energy and the parallel nonlinearity ${\cal N}_{\|}$.

\subsection{Energy density conservation at low speeds} 
In an inertial frame $O$,  let velocity fields $V_E$ and $V_B$ be 
associated with electric field $E$ and  magnetic field $B$, respectively. 
Then, the Lorentz force 
between a charged particle $q$, moving with velocity $V$,  and the fields is 
modeled as a 
function of the relative velocity between the particle and the field. 
The main idea is that a moving electric field introduces an apparent magnetic field 
(and vice versa); the model maintains a constant total energy density (electric and magnetic) 
that is independent 
of the relative velocity. This subsection begins with the relative-velocity-dependent 
modeling of the magnetic field.

\subsubsection{Low-speed, relative-velocity modeling of magnetic field}
%Relative-velocity-based Lorentz force due to magnetic field}
The Lorentz force on an electrical charge $q$ due to the magnetic field in terms of the 
relative velocity 
$V-V_B$ of the particle with respect to the field is 
\begin{equation} 
F_B ~~= ~q~(V-V_B) \times B
\label{Magnetic_force_on_Charge}
\end{equation}
Thus, the magnetic field $B$ appears to have an effective electric 
field $E_{B}$,  perpendicular to the relative velocity $V-V_B$, given by 
\begin{equation}
E_{B} ~~= ~(V-V_B) \times B
\label{E_field_due_to_B}
\end{equation}

This apparent electric field implies that the field energy would vary with 
the relative velocity of the charged particle in the same reference frame.  
To avoid such variation, a reduction of the apparent magnetic field $B_{B}$ 
in the perpendicular direction is considered so that the 
net energy of the apparent fields is independent of the relative velocity. 
In particular, 
it is assumed that the effective magnetic field (acting on an ideal magnetic 
particle that is moving with velocity $V$ according to observer $O$) 
is given by 

\begin{equation}
B_{B} ~~= ~B_{\|} ~+ \gamma_{B}B_{\bot}
\label{B_field_modification}
\end{equation}

\noindent 
where $\gamma_{B}B_{\bot}$ is the vector component of magnetic field 
perpendicular to the relative velocity $(V-V_B)$, and $ B_{\|}$ is the vector 
component of the magnetic field parallel to relative velocity $(V-V_B)$. 
In the nominal case, when the relative velocity is zero, i.e., $V=V_B$, 
we have no change in the perpendicular component and therefore, 
$\gamma_{B} =1$ for this case. When the relative velocity is nonzero,  
the factor $\gamma_{B}$ is chosen such that the net energy density of 
$B_{B}$ and $E_{B}$ (due to the magnetic field $B$) is independent of 
the relative velocity $(V-V_B)$.  Moreover, the only variations in the 
fields are in the perpendicular components.  
Therefore, by matching the energy density in the field's perpendicular 
component for the case when relative velocity is nonzero to the case 
when the relative velocity is zero, one obtains 
\begin{equation}
\frac{\gamma_{B}^2}{2\mu}|B_{\bot}|^2 ~+  
\frac{\epsilon}{2}
|E_{B}|^2
~=~ 
\frac{1}{2\mu}|B_{\bot}|^2
\label{Total_Energy}
\end{equation}
where  $ | \cdot |$ represents the magnitude of a vector,  $\epsilon$ 
is the permittivity,  and $\mu$ is the permeability. 
Substituting for the apparent electric field $E_{B}$ from 
equation~~(\ref{E_field_due_to_B}), i.e.,  
\begin{equation}
E_{B} ~= ~(V-V_B) \times B ~~ = (V-V_B) \times B_{\bot},
\end{equation}
into equation~(\ref{Total_Energy}), yields
\begin{equation}
\frac{\gamma_{B}^2}{2\mu}|B_{\bot}|^2 ~+  
\frac{\epsilon ~ |V-V_B|^2}{2}
|B_{\bot}|^2
~=~ 
\frac{1}{2\mu}|B_{\bot}|^2 
\label{Total_Energy_2}
\end{equation}
and   
\begin{equation}
\gamma_{B} ~ ~=~ \sqrt{{ {1} - \frac{|V-V_B|^2}{c^2} }} ~~ =~ 
\sqrt{{ {1} - \beta_B^2 }} ~~ ,  
\label{gamma_eq_B}
\end{equation}
where  $ c = {\sqrt{1/{\epsilon}{\mu}}} $ is the speed of light and 
$\beta_B$ is the normalized relative speed  
\begin{equation}
\beta_B = |V-V_B|/c . 
\label{magnetic_scaling_factor}
\end{equation} 
Thus, an electric particle moving with velocity $V$ is affected by 
the electric field $E_B$; a magnetic 
particle moving with velocity $V$ is affected by the magnetic field 
$B_B$, and the electric field $E_B$ moving 
with velocity $V_B$.

\subsubsection{Low-speed, relative-velocity modeling of electric field}
Similar to the last subsection, an electric 
field  $E$ appears to have an effective magnetic field $B_{E}$,  
perpendicular to the relative velocity 
$V-V_E$, given by 
\begin{equation}
B_{E} ~~= ~-{\epsilon}{\mu}(V-V_E) \times E
\label{B_field_due_to_E}
\end{equation}
where the term $-{\epsilon}{\mu}$ is used in 
equation~(\ref{B_field_due_to_E}) to match the 
magnetic field produced by a current-carrying wire (Ampere's law). 
In particular, 
if $\rho$ (charge per unit length) is flowing with velocity $v$ 
through a wire (which is stationary in the reference frame $O$)  
then 
the electric field $E_{\rho}$ associated with this charge, 
at a distance $r\hat{r}$ 
from the wire, is given by $ E_{\rho} = 
[\rho/(2{\pi\epsilon}r)]\hat{r} $, 
where $\hat{r}$ represents a unit direction vector. 
Note that the velocity associated with this electric 
field is the velocity $v$ of the charge flowing through the wire.  
Therefore, from equation~(\ref{B_field_due_to_E}), the magnetic field 
$B_{\rho}$ at a distance $r\hat{r}$ from the wire is 
\begin{equation}
\begin{array}{rcl}
B_{\rho} & =&  -({\epsilon}{\mu})~(0-v)\times E_{\rho} = 
({\epsilon}{\mu})~v \times E_{\rho}   \\
& = & {\epsilon}{\mu}[\rho/(2{\pi\epsilon}r)] |v|  ~~~~ 
\hat{v} \times \hat{r}  \\
& = & [\mu I/(2{\pi}r)] ~~~~ \hat{v} \times \hat{r} 
\end{array}
\end{equation}
where $I$ is the current in the wire; this is the  expression 
for magnetic field produced by a current-carrying wire. 

To keep the net energy independent of the relative velocity $V-V_E$, 
the following reduction $\gamma_{E}$ in the perpendicular direction of the apparent
electric field $E_{E}$  is considered 
\begin{equation}
E_{E} ~~= ~E_{\|} ~+ \gamma_{E}E_{\bot}
\label{E_field_modification}
\end{equation}
where $\gamma_{E}E_{\bot}$ is the vector component of electric field 
perpendicular to the relative velocity $(V-V_E)$, $ E_{\|}$ is the 
vector component of the electric field parallel to relative velocity 
$(V-V_E)$, and
the scaling factor $\gamma_{E} = 1$  when the relative velocity is 
zero, i.e., $V=V_E$. The scaling factor $\gamma_{E}$ is obtained 
by equating the total  energy density to the energy density of the 
electric field alone when the relative velocity is zero as   
\begin{equation}
{\frac{\epsilon\gamma_{E}^2}{2}}
{|E_{\bot}|^2} ~+  
{\frac{1}{2\mu}}
{|B_{E}|^2} ~=~ 
{\frac{\epsilon}{2}}
{|E_{\bot}|^2}. 
\label{Total_Energy_E}
\end{equation}
Substituting for the apparent magnetic field $B_{E}$ from 
equation~(\ref{B_field_due_to_E}), i.e.,  
\begin{equation}
B_{E} ~= ~-{\epsilon}{\mu}(V-V_E) \times E ~~ 
= ~-{\epsilon}{\mu}(V-V_E) \times E_{\bot}, 
\end{equation}
into equation~(\ref{Total_Energy_E}), yields
\begin{equation}
{\frac{\epsilon\gamma_{E}^2}{2}}
|E_{\bot}|^2 ~+  
{\frac{ \epsilon^2 \mu^2~ |V-V_E|^2 }{2\mu}}
|E_{\bot}|^2
~=~ 
\frac{\epsilon}{2}
|E_{\bot}|^2
\label{Total_Energy_22}
\end{equation}
and  a scaling factor 
\begin{equation}
\gamma_{E} ~ ~=~ \sqrt{{ {1} - \frac{|V-V_E|^2}{c^2} }} ~~ .   
\label{gamma_eq_E}
\end{equation}
This expression is similar to the one for the scaling factor for a 
magnetic field in equation~(\ref{gamma_eq_B}). 
Thus, a magnetic particle moving with velocity $V$ is affected by 
the magnetic field $B_E$; an electric 
particle moving with velocity $V$ is affected by the electric field 
$E_E$, and the magnetic field $B_E$ moving 
with velocity $V_E$. In particular, the net force on an electric 
particle (of charge $q$) is given by 
[from equations~(\ref{Magnetic_force_on_Charge},~\ref{B_field_due_to_E}) 
and~(\ref{E_field_modification})]
\begin{equation}
\begin{array}{rcl}
F_E  & = &  q~(V-V_E) \times B_E
~ + qE_{\|} ~+ q\gamma_{E}E_{\bot}    \\
& = &  q~(V-V_E) \times \left\{-{\epsilon}{\mu}(V-V_E) \times E  \right\} 
~~ + qE_{\|} ~+ q\gamma_{E}E_{\bot}   \\
& = &  q    \frac{|V-V_E|^2}{c^2}    E_{\bot}  
~ + qE_{\|} ~+ q E_{\bot} \sqrt{ {1} - \frac{|V-V_E|^2}{c^2}}
\\
& =  & q \left[  \beta_E^2 ~ + \sqrt{  {1} - \beta_E^2 }    
 \right]  E_{\bot} ~ + qE_{\|}   \\
& =  & q  \alpha  E_{\bot} ~ + q E_{\|}  
\label{electrical_field_force}
\end{array}
\end{equation}
where the normalized relative speed $\beta_E$  and the scaling factor 
$\alpha$ are given by 
\begin{equation}
\beta_E = |V-V_E|/c ~~ \mbox{and} ~~ \alpha = \beta_E^2 
~ + \sqrt{  {1} - \beta_E^2 } ~~ . 
\label{electrical_scaling_factor}
\end{equation}
When the relative velocity is small, i.e., $\beta_E$ is small, 
the scaling factor $\alpha$  
in  the perpendicular force component in equation~(\ref{electrical_field_force}) 
$$F_{E,\bot} ~=~ q \alpha  E_{\bot} $$ 
can be simplified to 
\begin{equation}
\alpha ~ = ~  \beta_E^2 ~ + \sqrt{  {1} - \beta_E^2 } 
~~~~~ \approx ~~ 1 + \frac{1}{2}\beta_E^2 ~.
\label{electrical_scaling_factor_simplified}
\end{equation}
Therefore, the simplified force on an electric particle in 
equation~(\ref{electrical_field_force}) 
becomes 
\begin{equation}
F_E  ~ \approx  ~  q \left( 1 + \frac{1}{2}\beta_E^2 \right) 
E_{\bot} ~ + q E_{\|}   
\label{electrical_field_force_simplified}
\end{equation}

\subsubsection{Saturation effect}
The discussions in the article are limited to the case 
when the magnitude of the 
relative velocity are less than the speed of light $c$, i.e., 
$\beta_E \le 1 $ and $\beta_B \le 1$. 
The approach can be extended to higher-relative speeds by 
fixing (saturating) the scaling factors to the values for 
the case when the relative speed equals the speed of light. 
For example,  $\alpha$ in equation~(\ref{electrical_scaling_factor}) 
is held constant for higher-relative speeds $\beta_E >1 $ as 
\begin{equation}
\alpha = 1  ~~ \forall ~ \beta_E > 1. 
\end{equation}
Although not stated explicitly,  equations are presented only for the case  
when $\beta_E \le 1 $ and $\beta_B \le 1$ in the rest of the article.

\subsection{High-speed effects in the relative-velocity model}
The relativistic mass dependence with speed is modeled as a slip effect, where the 
force on the particle reduces as the relative-velocity increases. In particular, 
consider the augmentation of the Lorentz force on an electric particle, 
in equations~(\ref{Magnetic_force_on_Charge}, ~\ref{electrical_field_force}), with  
relative-velocity terms $s_{\bot}$ and $s_{\|}$ as 
\begin{equation}
F_B  ~ = ~  \left[s_{\bot}(\beta_B)\right] q (V-V_B) \times B_{\bot}  
\label{Lorentz_force_modified_1} \\
\end{equation}
\begin{equation}
F_E  ~ = ~  \left[s_{\bot}(\beta_E)\right]  q \alpha  E_{\bot} ~ 
+ \left[s_{\|}(\beta_E)\right]  q E_{\|}  ~~~ = 
F_{E,\bot} ~~ + F_{E,\|} .  
\label{Lorentz_force_modified_2}
\end{equation}
The  perpendicular and parallel slip terms ($s_{\bot},s_{\|}$) 
are identified in this subsection.

\subsubsection{Matching cathode-ray-tube (CRT) observations}
Consider the forces on a charge moving with velocity $V$ perpendicular
to stationary magnetic $B$ and electric $E$ fields,  
as in  cathode-ray-tube (CRT) experiments (Thomson 1897). 
These forces can be written, 
from equations~(\ref{Lorentz_force_modified_1},~\ref{Lorentz_force_modified_2}),
as 
\begin{equation}
F_B ~= ~  ~ s_{\bot}(\beta) ~q~{V} \times B  \label{Slip_F_E_1} 
\end{equation}
\begin{equation}
F_E ~ = ~   
s_{\bot}(\beta)~ \alpha(\beta) ~q~E  \label{Slip_F_E_2} 
\end{equation}
where $\beta ~ = \frac{|{V}|}{c}$.    
If the fields act on the charged CRT particle over some length $L$, 
then the change in velocity of the CRT particle along the 
application of the force 
during the time interval $\Delta t = L/|V| $ is given by 
$$ \frac{F_B L}{m |V|} ~~{\mbox{and}} ~~\frac{F_E L}{m |V|} $$
where $m$ is the mass of the particle (electron). Therefore, 
the change in angles ($\theta_B$ and $\theta_E$) of the 
CRT particle's path along the action of the fields $B$ and 
$E$, respectively,  
can be approximated by using equations~(\ref{Slip_F_E_1},~\ref{Slip_F_E_2})
as~\cite{Thomson1897} 
\begin{equation}
\theta_B~~= ~ | \frac{F_B L}{m |V|^2} | ~= ~ 
\frac{s_{\bot}(\beta) ~q~{|V|} ~|B|~L }{m |V|^2}  ~~~ = ~ 
\frac{s_{\bot}(\beta)  ~|B|~L }{\frac{m}{q} |V|}  
\label{theta_slip_1} 
\end{equation}
\begin{equation}
\theta_E~~= ~ |\frac{F_E L}{m |V|^2}| ~ = ~   
\frac{s_{\bot}(\beta)~ \alpha(\beta) ~|E| ~L}{\frac{m}{q} |V|^2}  
\label{theta_slip_2}
\end{equation}

\vspace{0.1in}
\noindent
In the absence of the  relative-velocity terms (i.e., $s_{\bot}(\beta) =1 $ 
and~$\alpha(\beta) =1$), a velocity-dependent mass variation 
can be used to explain the CRT data. In particular, the estimated 
velocity $ V_{CRT}$ and the estimated mass-to-charge ratio 
$$ \frac{m (\beta_{CRT})}{q} = \frac{m}{q} ~\Psi(\beta_{CRT}),  $$
with
\begin{equation} \beta_{CRT} =  \frac{|{V_{CRT}}|}{c}, \end{equation}  
from the CRT experiments would be related by 
\begin{equation}
\theta_B ~ = ~  
\frac{    |B|~L }{\frac{m}{q} \Psi(\beta_{CRT}) |V_{CRT}|}  
\label{theta_no_slip_1} 
\end{equation}
\begin{equation}
\theta_E  ~ = ~   
\frac{  ~|E| ~L}{\frac{m}{q} \Psi(\beta_{CRT}) |V_{CRT}|^2}   
\label{theta_no_slip_2}
\end{equation}
where $\Psi(\beta_{CRT})$ represents the CRT-predicted 
variation of mass with velocity. 
Dividing equations~(\ref{theta_slip_1}) and~(\ref{theta_slip_2}) 
by equations~(\ref{theta_no_slip_1}) and~(\ref{theta_no_slip_2}), 
respectively, yields   
\begin{equation}
s_{\bot}(\beta) ~ = ~  \frac{|V|}{|V_{CRT}|} ~
\frac{ 1 }{  \Psi(\beta_{CRT})  }   
\label{theta_division_1}   
\end{equation}
\begin{equation}
s_{\bot}(\beta)~ \alpha(\beta) ~ = ~   \frac{|V|^2}{ |V_{CRT}|^2 } ~
\frac{  1 }{  \Psi(\beta_{CRT})    }.
\label{theta_division_2}    
\end{equation}
The  velocity $V_{CRT}$ predicted by the 
CRT-experiments can be obtained by dividing 
equation~(\ref{theta_division_1}) by equation~(\ref{theta_division_2}) 
to obtain 
\begin{equation} 
|V_{CRT}| =  \frac{|V|}{ \alpha(\beta) } ~~~ {\mbox{or}}~~  
\beta_{CRT} =  \frac{\beta}{ \alpha(\beta) }.  
\label{VCRT_VSLIP} 
\end{equation}
Furthermore, the perpendicular-slip term $s_{\bot}(\beta)$ can be 
found by dividing the square of equation~(\ref{theta_division_1}) 
by equation~(\ref{theta_division_2}) and then substituting for 
$\beta_{CRT}$ from equation~(\ref{VCRT_VSLIP}) to obtain 
\begin{equation} 
s_{\bot}(\beta) =  \frac{\alpha(\beta)}{\Psi(\beta_{CRT})} ~~= 
\frac{\alpha(\beta)}{\Psi(\frac{\beta}{ \alpha(\beta) })} .  
\label{nonlinearity} 
\end{equation}

\vspace{0.1in}
\noindent
{\underline{Case 1: matching the relativistic mass-velocity relation:~} } 
The perpendicular term $s_{\bot}(\beta)$ can be chosen, 
as in equation~(\ref{nonlinearity}), to exactly match the 
observed velocity-dependent variation 
$\Psi$ in mass. In particular, if the CRT-predicted mass 
increase is given by the relativistic expression   
\begin{equation}
\Psi(\beta_{CRT})  = \frac{1}{\sqrt{1 - \beta_{CRT}^2 } },  
\label{relativistic_mass_variation} 
\end{equation} 
then 
the expression for the slip term $s_{\bot}$ is obtained, 
from equation~(\ref{VCRT_VSLIP}) and equation~(\ref{nonlinearity}), as 
\begin{equation}
\begin{array}{rcl}
 s_{\bot}(\beta) & = & \alpha(\beta) ~~
\left\{ \sqrt{1 -  \left[ \frac{\beta }{ \alpha(\beta)  } \right]^2  
} \right\}     \\
& = &   \sqrt{ \left[\alpha(\beta)\right]^{2} -  \beta^2 }
\label{Slip_Case1} 
\end{array}
\end{equation}

\vspace{0.1in}
\noindent
{\underline{Case 2: simplified perpendicular slip term:~} } 
Consider the following,  simplified expression $\bar{s}_{\bot}$ for the slip term 
$s_{\bot}(\beta)$ 
\begin{equation}
\bar{s}_{\bot}(\beta) ~= ~ {\left[ {   1  ~-  \beta^8   } \right]^{1/4} }  
\label{s_approx}
\end{equation}
This term does not lead to an exact match of the relativistic mass 
increase; however, it closely approximates the expression for the 
relativistic mass increase. In particular, assuming this form 
$\bar{s}_{\bot}(\beta)$ for the slip term, 
the velocity $\bar{\beta}_{CRT}$ estimated in the CRT experiment, 
as in equation~(\ref{VCRT_VSLIP}),  
is  given by 
\begin{equation}
\bar{\beta}_{CRT} ~ = ~ \frac{\beta}{\alpha(\beta) } ~~ = ~
\frac{\beta}{\beta^2 +\sqrt{1 -\beta^2}}.
\label{Slip_Case2_1}
\end{equation}
Moreover, the  apparent  
mass variation $\bar{\Psi}$ in the CRT experiment, as in 
equation~(\ref{nonlinearity}),  is given by   
\begin{equation}
\bar{\Psi}(\bar{\beta}_{CRT})   ~ = ~ 
\frac{\alpha(\beta)}{s(\beta)}  ~~ = ~ 
\frac{\beta^2 +\sqrt{1 -\beta^2}}{\left[ {   
1  ~-  \beta^8   } \right]^{1/4}}
\label{Slip_Case2_3} . 
\end{equation}

\begin{figure}[!ht]
\begin{center}
\includegraphics*[width=3.25in]{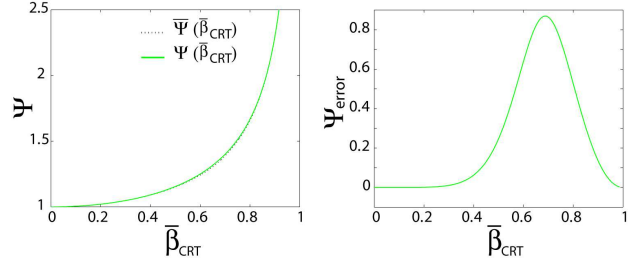}
\end{center}
\caption{Proposed model matches the apparent velocity dependence of mass in 
CRT experiments. Left plot: comparison of apparent mass variations ${\Psi}(\bar{\beta}_{CRT})$ 
(relativistic mass model) and $\bar{\Psi}(\bar{\beta}_{CRT})$ 
(simplified model) as in 
Eqs.~(\ref{Slip_Case2_3},~\ref{observed_mass_variation}) with normalized relative speed 
$\bar{\beta}_{CRT}$ in Eq.~(\ref{Slip_Case2_1}). Right plot: the difference $\Psi_{error}$ 
in predicted mass variation is less than 1\% with the simplified model in Eq.~(\ref{error}). 
An  match would have no error with the exact relative-velocity model in Eq.~(\ref{Slip_Case1}). 
}
\label{CRT_EXP}
\end{figure}

It is noted that the variation of $\bar{\Psi}(\bar{\beta}_{CRT})$ 
with velocity $\bar{\beta}_{CRT}$ in equations~(\ref{Slip_Case2_1}, \ref{Slip_Case2_3}), which would be obtained 
from a CRT experiment, is similar 
to the relativistic variation 
\begin{equation}
\Psi(\bar{\beta}_{CRT})  = \frac{1}{\sqrt{1 - \bar{\beta}_{CRT}^2 } }  
\label{observed_mass_variation} 
\end{equation}
as shown in figure~\ref{CRT_EXP}.

\vspace{0.1in}

Moreover, the percentage difference $\Psi_{error}$ between the two 
expressions (equations~\ref{Slip_Case2_3} and \ref{observed_mass_variation})
given by 
\begin{equation}
\Psi_{error} = \frac{{\Psi}(\bar{\beta}_{CRT}) 
- \bar{\Psi}(\bar{\beta}_{CRT})}{\Psi(\bar{\beta}_{CRT})} \times 100
\label{error} 
\end{equation}
is less than $1\%$ as shown in figure~\ref{CRT_EXP}. 
Thus, the relativistic velocity dependency 
of mass in CRT experiments can be modeled using  
the relative-velocity approach with the perpendicular nonlinearity  
(${\cal N}_{\bot} = s_{\bot} \alpha$) in the Lorentz force expression and a constant mass.  
The simplified expression for the perpendicular slip ($s_{\bot} = \bar{s}_{\bot}$ with 
$\bar{s}_{\bot}$ defined in Eq.~\ref{s_approx}) is used in the rest 
of the article.    

\subsection{Kinetic energy and parallel slip}
The expressions for kinetic energy ${{\cal{E}}_{KE}}$ and the parallel slip term 
${s}_{\|}$ are identified in this subsection by using the 
perpendicular slip term  ${s}_{\bot}$ . 

\subsubsection{Relationship between parallel slip and kinetic energy}
\begin{figure}[!ht]
\begin{center}
\includegraphics*[width=4in]{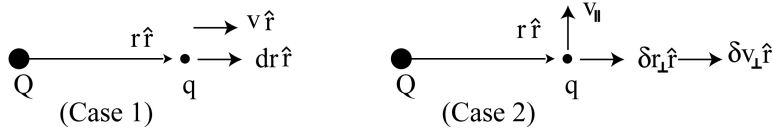}
\end{center}
\caption{Two cases: (i)~forces parallel; and (ii)~forces perpendicular to the  velocity.}
\label{fig:PE_KE}
\end{figure}

Consider a charged particle $q$ moving along a straight line away from a stationary charged 
particle $Q$ at a distance $r \hat{r}$ as shown in figure~\ref{fig:PE_KE} (case 1). 

Taking the dot product with a small displacement $dr \hat{r}$ with Newton's law 
on the charge $q$ yields 
\begin{equation}
s_{\|}(\beta) \frac{Qq}{4\pi\epsilon r^2} dr  ~~=~  
m \frac{dv}{dt} dr ~~ = \frac{m}{2} dv^2
~~ = \frac{mc^2}{2} d\beta^2. 
\label{Newtons_law_work_1} 
\end{equation}
Dividing both sides by the parallel slip term $s_{\|}(\beta)$ and integrating 
results in 
\begin{eqnarray}
\int_{r_1}^{r_2} \frac{Qq}{4\pi\epsilon r^2} dr  
& = & \int_{\beta^2_1}^{\beta^2_2}  \frac{mc^2}{2 s_{\|}(\beta)} d\beta^2 
~~ = \int_{{\cal{E}}_{KE,1}}^{{\cal{E}}_{KE,2}}  \frac{d{\cal{E}}_{KE}}{d\beta^2} d\beta^2 
\label{Newtons_law_work_2} 
\end{eqnarray}
where ${\cal{E}}_{KE}$ is considered as the relative-velocity dependent 
kinetic energy of the system since the above expression leads to the conservation law 
\begin{equation}
\frac{Qq}{4\pi\epsilon r_2} + {\cal{E}}_{KE,2} ~~ = \frac{Qq}{4\pi\epsilon r_1} + {\cal{E}}_{KE,1}
\label{Energy_conservation_E}
\end{equation}
in which the potential energy expression  ${Qq}/({4\pi\epsilon r})$ is independent of the relative velocity and the parallel slip term $s_{\|}(\beta)$. The relationship between the parallel slip term and the kinetic energy is  (from equation~\ref{Newtons_law_work_2}) 
\begin{equation}
 \frac{d{\cal{E}}_{KE}}{d\beta^2} ~~ = ~ \frac{mc^2}{2 s_{\|}(\beta)}. 
\label{Newtons_law_work_3} 
\end{equation}

\subsubsection{Expression for kinetic energy}
The perpendicular force $F_{E,\bot}$ does zero work and therefore does not lead to changes in the kinetic 
energy. However, an expression for kinetic energy can be found by 
making the virtual work done by the 
perpendicular force $F_{E,\bot}$ independent of the perpendicular slip term $s_{\bot}$. 
Consider a charged particle $q$ moving with velocity $v_{\|}$ perpendicular to the distance vector 
$r \hat{r}$ from a stationary charged particle $Q$ as shown in figure~\ref{fig:PE_KE} (case 2). 
Let $\delta{r_{\bot}} \hat{r}$ be a virtual displacement perpendicular to the relative velocity
$v_{\|}$; then taking the dot product of the virtual displacement with both sides of Newton's law
yields 
\begin{equation}
s_{\bot}(\beta_{\|})\alpha(\beta_{\|})\frac{Qq}{4\pi\epsilon r^2} \delta{r_{\bot}} ~~=~~ m \frac{\delta{v_{\bot}}}{\delta{t}}\delta{r_{\bot}} 
~~~  = m v_{\bot} \delta{v_{\bot}}  ~~~  = \frac{1}{2} m ~\delta{v_{\bot}^2}~ 
\label{V_work_1}
\end{equation}
where $\beta_{\|} = \frac{|v_{\|}|}{c}$. 
The virtual work done can be made independent of the 
slip term $s_{\bot}(\beta_{\|})$  if the 
change in the kinetic energy ($\delta{{\cal{E}}_{KE}}$) 
has the following form (obtained by dividing both sides of the above equation by 
$s_{\bot}(\beta_{\|})$)
%$\frac{1}{2 s_{\bot}(\beta_{\|})}m ~\delta v_{\bot}^2  $, i.e., 
\begin{equation}
\alpha(\beta_{\|})\frac{Qq}{4\pi\epsilon r^2} \delta{r_{\bot}} ~~=~~  \frac{1}{2} 
\frac{ m ~\delta{v_{\bot}^2}}{s_{\bot}(\beta_{\|})} 
~~=~~  \frac{1}{2} \frac{ m c^2 ~\delta{\beta_{\bot}^2}}{s_{\bot}(\beta_{\|})} 
~~ = ~ \delta{{\cal{E}}_{KE}}
\label{V_work_2}
\end{equation}
which implies that the  kinetic energy ${{\cal{E}}_{KE}}$ has the form 
\begin{equation}
{{\cal{E}}_{KE}}(\beta)~~ =~  \frac{1}{2} \frac{m c^2 \beta^2}{s_{\bot}} ~~~~~= \frac{1}{2} \frac{m c^2 \beta^2}{(1 -\beta^8)^{1/4}} . 
\label{Virtual_work} 
\end{equation}

\subsubsection{Expression for parallel slip term}
Differentiating the expression (equation~\ref{Virtual_work}) for the kinetic energy by $\beta^2$ 
yields 
\begin{equation}
 \frac{d{\cal{E}}_{KE}}{d\beta^2} ~~ 
= ~ \frac{mc^2}{2} \left[ 
\frac{1}{(1 -\beta^8)^{5/4}}
\right]. 
\label{derivative_E_KE} 
\end{equation}
and comparison with equation~(\ref{Newtons_law_work_3}) yields the parallel slip term as 
\begin{equation}
 s_{\|} ~~ 
= ~ (1 -\beta^8)^{5/4}. 
\label{parallel_slip_term} 
\end{equation}

\subsection{Summary of relative-velocity-dependent model}
The relative velocity approach results in the following Lorentz force on an electrically charged particle 
(from equation~\ref{Lorentz_force_modified_1} and equation~\ref{Lorentz_force_modified_2}) 
by using the simplified slip terms 
(in equation~\ref{s_approx} and equation~\ref{parallel_slip_term}) 

\begin{equation}
F_B  ~ = ~  \left[ (1 -\beta_B^8)^{1/4} \right] q (V-V_B) \times B_{\bot}  
\label{Lorentz_force_modified_3} 
\end{equation}
\begin{equation}
F_E  ~ = ~  \left[(1 -\beta_E^8)^{1/4}  \right]  \left[\sqrt{1 -\beta_E^2} + \beta_E^2 \right] q   E_{\bot} ~ 
+ \left[(1 -\beta_E^8)^{5/4}\right]  q E_{\|} . 
\label{Lorentz_force_modified_4}
\end{equation}
where the electrical force expression (in equation~\ref{Lorentz_force_modified_4}) reduces to the expression in equation~(\ref{electrical_field_force_simplified}) at low speeds.

%\newpage
\section{Applications of Lorentz force expression}
In this section, it is shown that relative-velocity-dependent Lorentz force 
expression: (a)~satisfies the force between two wires (a low-velocity  effect); and (b)~explains the 
observed discrepancy in the energy of high-velocity particles. 
Moreover, it is shown that similar relative-velocity dependence of the gravitational force can explain   
the excess precession in planetary orbits.  

\subsection{Force between two wires}
\begin{figure}[!ht]
\begin{center}
\includegraphics*[width=1.2in]{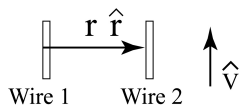}
\end{center}
\caption{Force between two current carrying parallel wires 
separated by distance $r$.}
\label{fig:Parallel_wires}
\end{figure}

The increase in the electrical force component perpendicular to the relative 
velocity [in equation~(\ref{Lorentz_force_modified_4}) which simplifies to  equation~(\ref{electrical_field_force_simplified}) at low speeds]
can be used to explain the force between two current carrying wires, 
which are both stationary in a reference frame $O$. 
As shown in figure~\ref{fig:Parallel_wires}, let the second wire (denoted by the subscript 2) be 
positioned at $r\hat{r}$ from the first wire (denoted by the subscript 1). 
Moreover,  
let the currents in the two parallel wires be  $I_1$ and $I_2$, and 
let the corresponding moving charges (per unit length) be $-\rho_1$ 
and $-\rho_2$ with 
velocities $-v_1\hat{V}$ and $-v_2\hat{V}$, respectively, i.e., 
\begin{equation} 
I_1 = \rho_1 v_1, ~{\mbox{and}} ~
I_2 = \rho_2 v_2
\end{equation}
where the speeds $v_1 \ge 0$ and $v_2 \ge 0$ of the charges are small, 
and $\hat{V}$ is a unit vector along the direction of the wire 
(in which current is flowing).  

\subsubsection{Force expression}
Consider the two (non-canceling) fields in the first 
wire:~(a)~$E_{-\rho_1}$  associated with the moving charge $-\rho_1$ 
with field velocity $-v_1\hat{V}$ given by 
\begin{equation} 
E_{-\rho_1} = [-\rho_1/(2{\pi\epsilon}r)]\hat{r};
\label{wire_field_minus}
\end{equation}
and 
(b)~$E_{\rho_1}$ associated with the corresponding stationary charge $\rho_1$ 
in the wire, i.e.,  
the stationary field given by 
\begin{equation} 
E_{\rho_1} = [\rho_1/(2{\pi\epsilon}r)]\hat{r}. 
\label{wire_field_plus}
\end{equation}
These two fields act on the moving charge $-\rho_2$ and a corresponding 
stationary charge $\rho_2$ on the second wire (per unit length). 
For example, the force per unit length $F_{-\rho_1,-\rho_2} $
on the moving charge  $-\rho_2$ due to the moving charge $-\rho_1$ 
can be obtained from 
equations~(\ref{electrical_field_force_simplified},~\ref{wire_field_minus}) 
as 
\begin{equation}
\begin{array}{rcl}
F_{-\rho_1,-\rho_2}  & =  &  -\rho_2 \left( 1 + \frac{|(-v_2) 
-(-v_1) |^2}{2c^2} \right) E_{-\rho_1}  \\
& = &  -\rho_2 \left( 1 + \frac{|(-v_2) -(-v_1) |^2}{2c^2} \right) 
[-\rho_1/(2{\pi\epsilon}r)]\hat{r} \\
& = & 
\frac{\rho_1 \rho_2 \hat{r}}{2{\pi\epsilon}r} 
\left( 1 + \frac{|v_1 -v_2|^2}{2c^2} \right)  . 
\label{F_wire_mm}
\end{array}
\end{equation}
Similarly, (i)~the force per unit length $F_{\rho_1,-\rho_2} $
on the moving charge  $-\rho_2$ due to the stationary charge 
$\rho_1$, as well as (ii)~the 
the forces $F_{-\rho_1,\rho_2} ,F_{\rho_1,\rho_2} $  on the 
stationary charge  $\rho_2$ on the second wire due to the charges (on the first wire) 
$-\rho_1$ and $\rho_1$, respectively, are given by 
\begin{equation}
\begin{array}{rcl}
F_{\rho_1,-\rho_2} & = &  -\frac{\rho_1 \rho_2 \hat{r}}{2{\pi\epsilon}r} 
\left( 1 + \frac{|-v_2|^2}{2c^2} \right)  \\
F_{-\rho_1,\rho_2} & = &  -\frac{\rho_1 \rho_2 \hat{r}}{2{\pi\epsilon}r} 
\left( 1 + \frac{|v_1|^2}{2c^2} \right) 
\label{F_wire_other} \\
F_{\rho_1,\rho_2} & = &  \frac{\rho_1 \rho_2 \hat{r}}{2{\pi\epsilon}r}  ~.
\end{array}
\end{equation}
Thus, the total force per unit length $F_{1,2}$ on the second wire can be 
found using equations~(\ref{F_wire_mm},~\ref{F_wire_other}) as 
\begin{equation}
\begin{array}{rcl}
F_{1,2}  & =  & F_{\rho_1,\rho_2} + F_{\rho_1,-\rho_2} +  F_{-\rho_1,\rho_2}
+  F_{-\rho_1,-\rho_2}   \\
& =  & \frac{\rho_1 \rho_2 \hat{r}}{2{\pi\epsilon}r}  \left[{  
1-\left( 1 + \frac{|v_2|^2}{2c^2} \right)
-\left( 1 + \frac{|v_1|^2}{2c^2} \right)
+\left( 1 + \frac{|v_1 -v_2|^2}{2c^2} \right)
}\right]  \\
& =  & \frac{\rho_1 \rho_2 \hat{r}}{2{\pi\epsilon}r}  ~ 
\left(\frac{-2v_1v_2}{2c^2}\right) ~~
= ~ - \frac{\mu I_1 I_2 }{2{\pi}r} \hat{r}. 
\label{F_12}
\end{array}
\end{equation}
This force on the second wire is attractive (i.e., towards the first wire) 
when the two wires carry 
current in the same direction.

\subsubsection{Force between wires is incorrect with ad-hoc perpendicular slip term}
Another choice of the perpendicular slip term can be found by matching the 
acceleration resulting from the 
relativistic increase in mass with speed. 
For example, the slip term $s_{\bot}(\beta_E)$ in 
equation~(\ref{Lorentz_force_modified_2}) can be chosen such 
that the perpendicular component of the force due to an 
electric field becomes 
\begin{equation}
F_{E,\bot}  ~ = ~~  \left[s_{\bot}(\beta_E)\right]  
q \alpha  E_{\bot} ~~ =    q  E_{\bot} \sqrt{1 - \beta_E^2}. 
\label{Lorentz_force_relativity_full}
\end{equation}
The resulting force can be approximated (at low speeds $\beta_E$) by 
\begin{equation}
F_{E,\bot}  ~ \cong ~~     q  E_{\bot} 
\left( 1 - \frac{\beta_E^2}{2} \right) .
\label{Lorentz_force_relativity}
\end{equation} 
However, this expression is not consistent with the 
force between two current carrying wires. In particular, the 
scaling of the term $\beta_E^2$ has the opposite sign in 
equation~(\ref{Lorentz_force_relativity}) when compared to the 
expression in equation~(\ref{electrical_field_force_simplified}). 
The use of the expression in equation~(\ref{Lorentz_force_relativity}) 
would lead to a force 
\begin{equation}
F_{1,2}  ~ =  \frac{\mu I_1 I_2 }{2{\pi}r} \hat{r}
\end{equation} 
similar to equation~(\ref{F_12}); however, the force between two 
wires that are carrying current in the same direction is 
repulsive, which is incorrect. 
%
%The need to match  the force 
%between two current carrying wires indicates that the perpendicular slip term 
%should be either constants or include terms higher than $\beta_E^2$.  
%

\subsection{High-velocity particles} 
The energy and velocity of charged particles in magnetic fields predicted by 
the relative velocity approach are compared with relativistic predictions. 
These are then used to explain apparent discrepancies in the measured energy of 
high-velocity experiments. 

\subsubsection{Comparison of energy and velocity expressions}
\begin{figure}[!ht]
\begin{center}
\includegraphics*[width=3.25in]{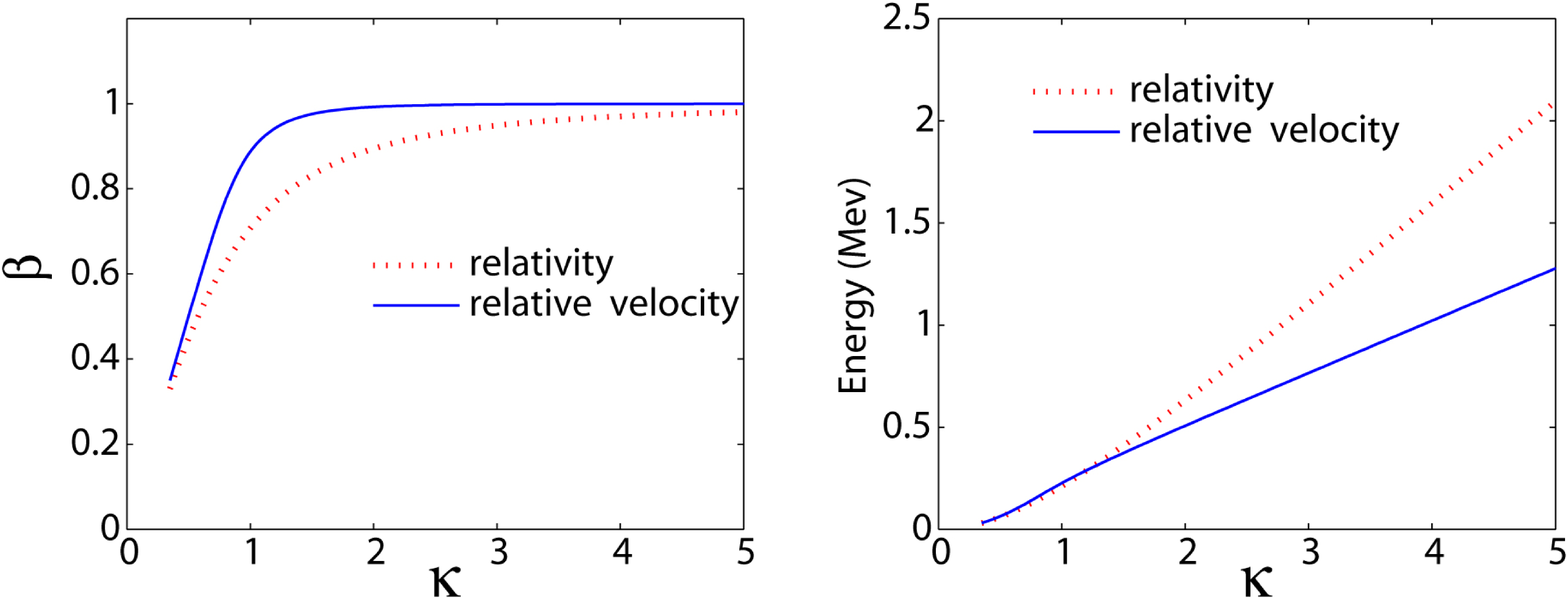}
\end{center}
\caption{Variation of normalized speed (left plot) and energy (right plot).}
\label{Fig_4_energy_beta_comparison}
\end{figure}

The motion of a particle with charge $q$ and mass $m$ moving with speed $v$ 
in a magnetic field $B$ along a circle with radius $\rho$ is governed by 
\begin{equation}
\frac{m v^2}{\rho} ~ = ~   q v s_{\bot} B ~~~~~ {\mbox{i.e.,}}~~
 \frac{\beta}{(1 - \beta^8)^{1/4} }  ~~ =~ \left(\frac{q}{m c}\right) \rho B  ~~ = \kappa. 
\label{rv_vel}
\end{equation}
The relativistic expression for velocity $v_r$ (and $\beta_r = |v_r|/c$), for a given value of the 
non-dimensional parameter $\kappa$, 
can be found as 
\begin{equation}
\frac{m v_{r}^2}{\rho \sqrt{1 - \beta_{r}^2} } ~ = ~   q v_r  B ~~~~~ {\mbox{i.e.,}}~~
 \frac{\beta_r}{(1 - \beta_r^2)^{1/2} }   ~~ = \kappa 
\label{relativity_beta}
\end{equation}
where $m$ represents the rest mass in relativity. The corresponding relativistic energy ${\cal{E}}_r$ is given by 
\begin{equation}
{{\cal{E}}_{r}}  ~ = ~  \frac{m c^2}{(1 - \beta_{r}^2)^{1/2} } -m c^2   
\label{relativity_E}
\end{equation}
while the relative-velocity-based energy ${\cal{E}}$ is given by equation~(\ref{Virtual_work}). 
The  predicted velocity and energy for the relative velocity approach are compared with those from the relativistic expressions in figure~\ref{Fig_4_energy_beta_comparison}. 

Note that the  velocity $\beta$ predicted by the relative velocity approach tends to be higher than the relativistic value $\beta_r$ while the energy ${\cal{E}}$ predicted by the relative velocity approach tends to be lower than the relativistic value ${{\cal{E}}_{r}}$ for large values of $\kappa$. 
This difference is used to explain (below) the discrepancies in the 
observed energy in absorption experiments and in the 
$\beta$-ray spectrum.

\subsubsection{Absorption of high-energy electrons}
\begin{figure}[!ht]
\begin{center}
\includegraphics*[width=3.25in]{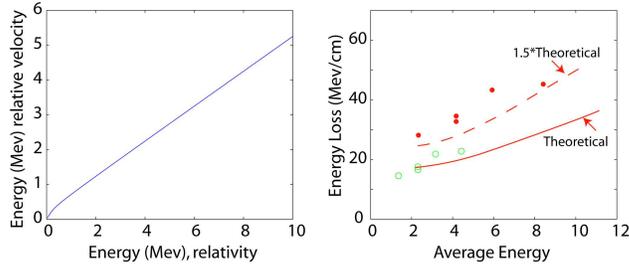}
\end{center}
\caption{Relative-velocity-based energy vs. relativistic energy  
in magnetic fields (left plot). 
Right plot:   the large deviation between predicted and experimental energy loss $\Delta {{\cal{E}}}$ for different average  energy ${\bar{\cal{E}}}$
is avoided  with the relative-velocity-based approach. 
The relative-velocity-based values (circles) are from Table~\ref{table_crane};  the
relativity-based values (solid dots) and the 
theoretical curves (solid/dashed lines) are from Turin and Crane~\cite{Turin_Crane_II}, figure~7.}
\label{Fig_5_crane_turin_data}
\end{figure}

Crane and co-workers~\cite{Turin_Crane_II} investigated the absorption of high-energy electrons in lead by measuring the 
initial energy ${{\cal{E}}_{i,r}}$ and final energy ${{\cal{E}}_{f,r}}$  
of electrons before and after passing through a  
lead absorber in a cloud chamber.~\cite{Turin_Crane_II} \  
The subscript $r$ denotes that the relativistic expression is used to find the energy
from the measured curvature in a magnetic field. The corresponding initial and final 
energy, from the relative-velocity approach, are found from figure~\ref{Fig_5_crane_turin_data}~plot(a)  
which plots the energy from the relative-velocity approach against the corresponding 
relativistic energy at different radii in the magnetic field (i.e., different $\kappa$). 
Data from Crane's results (Turin and Crane,~\cite{Turin_Crane_II} \ figure~7) is used to recompute the energy loss using the 
relative velocity approach --- the results are shown in Table~\ref{table_crane} 
and in figure~\ref{Fig_5_crane_turin_data}~plot(b) which plots the energy loss 
$ \Delta {{\cal{E}}} = {{\cal{E}}_i} - {{\cal{E}}_f}$ versus the 
average energy 
$  {\bar{\cal{E}}}  = ({{\cal{E}}_i} - {{\cal{E}}_f})/2$.

\begin{table}
\begin{center}
\vspace{0.2in}
\begin{tabular}{|c|c| c| c|c |c |c|c|}
% after \\: \hline or \cline{col1-col2} \cline{col3-col4} ...
\hline
\multicolumn{4}{|c|}{Relativity approach} &\multicolumn{4}{c|}{Relative velocity }\\
\cline{1-8}
average & loss & initial & final & initial & final  & average & loss   \\
${\bar{\cal{E}}_{r}}$ & $\Delta {{\cal{E}}_r}$  & $ {{\cal{E}}_{r,i}}$ & ${{\cal{E}}_{r,f}}$ & 
$ {{\cal{E}}_i}$     & $ {{\cal{E}}_f}   $ & $ {\bar{\cal{E}}}   $ & $\Delta {{\cal{E}}}$     
\\
Mev & Mev/cm & Mev & Mev & Mev & Mev & Mev & Mev/cm  \\
\hline\hline 
2.27  &  28.5  & 2.983  &  1.558   & 1.728  & 1.002 & 1.365 &  14.424 \\
4.12  &  33    & 4.945  &  3.295   & 2.716  & 1.886 & 2.301 &  16.606 \\
4.12  &  35    & 4.995  &  3.245   & 2.741  & 1.861 & 2.301 &  17.613 \\
5.85  &  43.5  & 6.938  &  4.763   & 3.716  & 2.624 & 3.170 &  21.823 \\
8.35  &  45.5  & 9.488  &  7.213   & 4.993  & 3.853 & 4.423 &  22.789 \\
\hline
\end{tabular}
\end{center}
\caption{Re-evaluation of energy loss in electron absorption from data by Turin and Crane.~\cite{Turin_Crane_II} \  Thickness of the lead absorber is $0.5mm$.~\cite{Turin_Crane_II}}
\label{table_crane}
\end{table}

Thus, the results show that the  large deviation (more than 50\% increase) in the energy loss from the theoretical prediction (Turin and Crane,~\cite{Turin_Crane_II} \ figure~7)  when energy is computed with the relativistic 
approach is avoided when the data is re-interpreted with the relative-velocity-based expression for 
energy.

%\newpage
\subsubsection{Average energy of $\beta$-ray spectrum}
\begin{figure}[!ht]
\begin{center}
\includegraphics*[width=3.25in]{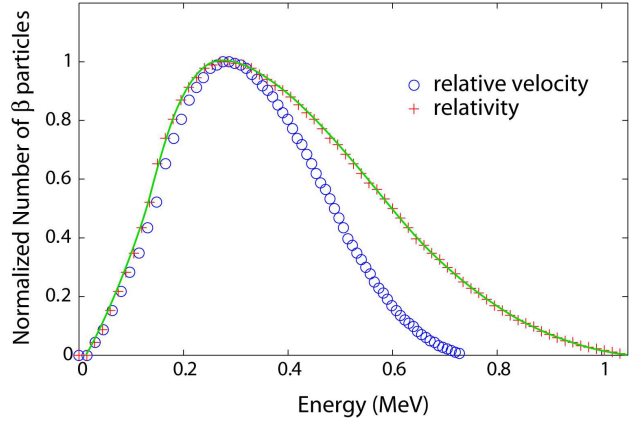}
\end{center}
\caption{$\beta$-particle counts vs energy. The relative-velocity approach predicts a reduction in the number of 
particles at the high-energy end of the spectrum; the resulting average energy is closer to the temperature-based measurement by Ellis and Wooster.~\cite{Ellis_Wooster_1927} \ The solid line is Madgwick's curve (from Ellis and Wooster,~\cite{Ellis_Wooster_1927} \  Fig.~1) with the relativistic energy expression. The crosses are data points obtained from the solid curve; these data points are used to find the corresponding relative-velocity-based energy (circles).}
\label{Fig_6_Ellis_Wooster}
\end{figure}

Ellis and Wooster~\cite{Ellis_Wooster_1927} 
found the average disintegration energy of 
Radium~E (Ra~E) to be $0.344MeV$ using temperature measurements while 
the  average energy found from the $\beta$-ray spectrum in magnetic fields 
(Ellis and Wooster~\cite{Ellis_Wooster_1927},  figure~1) 
was $0.395MeV$. 
The resulting spectrum~\cite{Ellis_Wooster_1927} 
is shown in figure~\ref{Fig_6_Ellis_Wooster} (solid line); data points
were measured on this curve  and the corresponding relative-velocity based energy was found using the same relationship in figure~\ref{Fig_5_crane_turin_data}(plot a) --- the recomputed data points are also shown 
in figure~\ref{Fig_6_Ellis_Wooster}.

%\newpage
Note 
that the value of the energy is smaller for the relative-velocity approach when compared to the relativistic energy (towards the higher end of the spectrum in figure~\ref{Fig_6_Ellis_Wooster}); therefore,  
the average energy is smaller with the relative-velocity approach.  
The average value of the spectrum with the relativistic data points is $0.39 Mev$ while the average value of the spectrum with the relative-velocity-based points is $0.35 Mev $. Thus, the  re-computed average energy ($0.35 Mev$) of Madgwick's $\beta$-ray spectrum data is close to the average ($0.344 Mev$)  
obtained by Ellis and Wooster~\cite{Ellis_Wooster_1927} using temperature measurements when compared to the value of $0.39 Mev$ with the relativistic energy expression~\cite{Ellis_Wooster_1927}. Therefore,  the apparent 
discrepancy in  Madgwick's data~\cite{Ellis_Wooster_1927} can be  explained by using the relative-velocity-based approach.

%\subsubsection{Lifetime of Mesons}

\subsection{Precession of planetary orbits}
It is shown that relative-velocity 
dependence of the normal force  can be used to explain 
the additional precession in planetary orbits. 

\subsubsection{Gravitational force expression with slip terms}
Consider a similar nonlinearity (as in equation~\ref{Lorentz_force_modified_4}) for modeling the 
gravitational force on a planet of mass $m_p$ 
\begin{eqnarray}
F_G & = &  
s_{\bot}(\beta)  \left[ \alpha_G(\beta) \right]  m_p  G_{\bot}  ~ 
+ s_{\|}(\beta)    m_p G_{\|}  
~~~ = F_{G,\bot} ~ +  F_{G,\|}
\label{nonlinear_G} 
\end{eqnarray}
where $G_{\|}$ and $G_{\bot}$ represents the gravitational field components 
due to the sun (mass $M_s$) that are parallel and perpendicular to the relative 
velocity $v_{rel}$ with respect to the gravitational field and  $\beta = |v_{rel}|/c$.  
If the expression for kinetic energy is kept the same for  
gravitational fields as with electrical or magnetic fields, then the slip terms ($s_{\bot}, s_{\|}$)
will be the same for gravitational fields. However, there is still flexibility in the choice 
of the term $\alpha_G(\beta) $ in the normal force. 

\subsubsection{Perturbation of the potential}
For planetary orbits, the speeds are small (therefore $\beta^8$ is small and the slip terms are close to one), and the orbits are almost circular hence the central force is close to the normal force $F_{G,\bot}$. 
Therefore,  the gravitational  
force on the planet (in equation~\ref{nonlinear_G}) at a distance $r \hat{r}$ from the sun can be approximated by the normal force given by 
\begin{eqnarray}
F_G & = &  -  G \frac{m_p M_s}{r^2}  \left[ \alpha_G(\beta) \right] \hat{r} 
~~ =   -  G \frac{m_p M_s}{r^2}  \left[ 1+ K_G {\beta^2} \right] \hat{r}   
\label{normal_force_G} 
\end{eqnarray}
where $G$ is the gravitational force constant, $m_p$ is the mass of the planet, $M_s$ is the mass of the sun,  and $K_G$ is a constant. Energy conservation 
(obtained using the same integration procedure used to obtain equation~\ref{Energy_conservation_E}) with the kinetic energy 
expression in equation~(\ref{Virtual_work}) at low speeds yields 
\begin{eqnarray}
-  G \frac{m_p M_s}{r} + \frac{1}{2} \frac{m_p c^2 \beta^2}{(1 -\beta^8)^{1/4}} 
~~ \approx ~ 
-  G \frac{m_p M_s}{r} + \frac{1}{2}{m_p c^2 \beta^2}
~= {\cal{E}}
\label{Energy_G} 
\end{eqnarray}
where ${\cal{E}}$ is a constant --- the total energy at any point on the orbit. 
Substituting for 
$
\beta^2 = \frac{2\cal{E}}{m_p c^2}  + G \frac{2 M_s}{r c^2}  
$
from equation~(\ref{Energy_G}) into equation~(\ref{normal_force_G}) results in 
a force expression 
\begin{eqnarray}
F_G & = &   
-  G \frac{m_p M_s (1+ \frac{2K_G\cal{E}}{m_p c^2})}{r^2}  \hat{r}  
~ 
-  2K_G G^2 \frac{m_p M_s^2}{c^2 r^3}  \hat{r}  
\label{normal_force_G_2} 
\end{eqnarray}
and an associated potential-like function $V(r)$ such that $F_G(r)  = - \nabla V(r) $ given by 
\begin{eqnarray}
V & = &   
-  \frac{G m_p M_s (1+ \frac{2K_G\cal{E}}{m_p c^2})}{r}  
~ 
-  \frac{K_G G^2 m_p M_s^2 }{c^2 r^2}   \nonumber \\
& = &   
-  \frac{G m_p M_s (1+ \frac{2K_G\cal{E}}{m_p c^2})}{r}  
~ 
-  \frac{h}{ r^2} 
\label{potential_G} 
\end{eqnarray}
where $ h = \frac{K_G  G^2 m_p M_s^2 }{c^2}$ and  the $1/r$ potential is perturbed by 
$-{h}/{ r^2} $. 

\subsubsection{Precession due to  perturbation of the potential}
The precession rate $\dot{P}$ for a  perturbation  of the potential ($V$) with the form 
$-h/r^2$ is equal to (see Goldstein,~\cite{Goldstein_80} \ equations~(11-46 and 11-50))
\begin{equation}
\dot{P} = \frac{2\pi  h }{ G M_s m_p a ( 1 - \epsilon ^2) T_p} 
~~ = ~  \frac{2 \pi K_G  G  M_s }{  a ( 1 - \epsilon ^2) T_p c^2}  
\end{equation}
where $T_p$ is orbital time period, $a$ 
is the semi-major axis and $\epsilon$ is the eccentricity of the orbit. 
This matches the general relativistic prediction of precession rate given by 
(see Goldstein,~\cite{Goldstein_80} \ equations~(11-51 and 11-52)) 
\begin{equation}
\dot{P}~ = ~  \frac{6 \pi  G  M_s }{  a ( 1 - \epsilon ^2) T_p c^2}  
\end{equation}
when the constant $ K_G =3$.  
It is noted that the term $(1 + \frac{2K_G\cal{E}}{m_p c^2})$ is approximately one 
in equations~(\ref{normal_force_G_2}, \ref{potential_G}) since $ \frac{2K_G\cal{E}}{m_p c^2}$
is of the order $10^{-8}$ and 
is negligible. 
The resulting gravitational force expression is 
\begin{eqnarray}
F_G & = &  
\left[(1 -\beta^8)^{1/4}  \right]  \left[ 1 + 3\beta^2 \right]  m_p  G_{\bot}  ~ 
+ \left[(1 -\beta^8)^{5/4}\right]   m_p G_{\|} 
\label{final_G} 
\end{eqnarray}

Thus, the relative-velocity dependent approach can predict the excess precession 
of planetary orbits. 

\section{Optics} 
The field velocities ($V_E, V_B$) introduce  extra terms in Maxwell's equations
that are removed to retain co-ordinate invariance. It is shown that the proposed model 
captures relativistic effects in: 
(i)~the 
propagation speed of light; (ii)~stellar aberration;  
(iii)~the transverse Doppler effect; and 
(iv)~the convection of light by moving media.  

\subsection{Relative-velocity in Maxwell's equations}
Consider an electric $E$ and a magnetic $M$ field, which are 
stationary with respect to an inertial frame $O_1$, and 
satisfy Maxwell's equations 
in free space without charges 
\begin{equation}
\nabla \times E  ~ = ~  - \frac{{\partial}B}{{\partial}t} 
\label{Faraday_Eq_nom} 
\end{equation}
\begin{equation}
\nabla \times B  ~ = ~   \epsilon  \mu \frac{{\partial}E}{{\partial}t}. 
\label{Ampere_Eq_nom}
\end{equation}
Consider the same equation in a different inertial frame $O_2$ in 
which the inertial frame $O_1$ is moving with constant velocity $V$. 
The Galilean transformation between the two frames $$X_2 = X_1 + Vt$$  
gives the following relations  
at any location ($X_1$ in frame $O_1$ or $X_2-Vt$ in frame $O_2$)  
%\begin{equation}  
\begin{eqnarray}
{\mbox{Frame}}~O_1  & ~~~~~~~ & {\mbox{Frame}}~O_2   \\
E(a,b), ~~B(a,b) & & E(a,b), ~~B(a,b)\\
a = X_1, ~~b=t, & & a = X_2-Vt, ~~b=t \\
V_E = 0, ~~V_B = 0 & & V_E = V, ~~V_B = V  \\
\frac{{\partial}E}{{\partial}t} = \frac{{\partial}E}{{\partial}b} & & 
\frac{{\partial}E}{{\partial}t} = \frac{{\partial}E}{{\partial}a}(-V_E) 
~+\frac{{\partial}E}{{\partial}b} \\
& & 
~~~~~  =    -(V_E\cdot\nabla)E~+\frac{{\partial}E}{{\partial}b}  
\label{time_gradient} \\
\frac{{\partial}E}{{\partial}X_1} = \frac{{\partial}E}{{\partial}a} 
& & 
\frac{{\partial}E}{{\partial}X_2} = \frac{{\partial}E}{{\partial}a}.   
\label{spatial_gradient}
\end{eqnarray}
%\end{equation}
Since the spatial gradient is invariant with frame, 
in equation~(\ref{spatial_gradient}), the curl --- e.g., $\nabla \times B$ 
on the left hand side of equation~(\ref{Ampere_Eq_nom}) --- is also frame 
invariant. However, the partial time derivative 
in equation~(\ref{time_gradient}) has an extra term in frame 2. Therefore, 
the partial derivative with time, such as $\frac{{\partial}E}{{\partial}t}$ 
on the right hand side of equation~(\ref{Ampere_Eq_nom}), has an extra 
term $-(V_E\cdot\nabla)E$. Hence, adding the term $(V_E\cdot\nabla)E$ 
to Maxwell's equation~(\ref{Ampere_Eq_nom}) will make it frame invariant under the 
relative-velocity-dependent approach; 
the modified equation becomes 
\begin{equation}
\nabla \times B  ~ = ~   \epsilon  \mu 
\left( 
\frac{{\partial}E}{{\partial}t} 
~+(V_E\cdot\nabla)E 
\right). 
\label{Ampere_Eq_temp}
\end{equation}
Noting that 
\begin{equation} \frac{dE}{dt} ~=~
\frac{{\partial}E}{{\partial}t} 
~+(V_E\cdot\nabla)E 
\end{equation}
and using a similar argument to modify equation~(\ref{Faraday_Eq_nom}), 
we obtain the following inertial-frame invariant form of Maxwell's 
equations with terms that include the field velocities ($V_E, V_B$)
\begin{equation}
\nabla \times E  ~ = ~  - \frac{dB}{dt} 
\label{Faraday_Eq} 
\end{equation}
\begin{equation}
\nabla \times B  ~ = ~  \epsilon  \mu \frac{dE}{dt}   
\label{Ampere_Eq}
\end{equation}

\subsubsection{Invariance with co-ordinate change}
Electric $E$ and magnetic $B$ fields, with field velocities 
$V_E$ and  $V_B$, respectively, that satisfy Maxwell's 
equations in one reference frame 
also satisfy it in another inertial reference frame with 
a Galilean transformation of the field velocities. 
In this sense, the modified Maxwell's 
equations [equations~(\ref{Faraday_Eq}),~(\ref{Ampere_Eq})] with the total 
time derivatives are invariant to Galilean transformations 
between inertial reference frames. 

\subsubsection{Addition of current density}
It is noted that a current density of the form 
\begin{equation} \mu J = \mu \epsilon \left({  
\nabla \cdot E }\right) V_E 
\end{equation} 
can be added to 
the right hand side of equation~(\ref{Ampere_Eq}) but is not needed 
in the following discussion on optics.

\subsection{Propagation Speed of Light} 
Consider the following two wave equations, which are considered 
as disturbances on the nominal 
electrical $E$ and magnetic $B$ fields, each of which has a 
field velocity $$V_E = V_B = V = v_z \hat{z},$$ with magnitude 
$v_z$ in the $\hat{z}$ direction, given by   
\begin{equation}
E ~=~ e_x \cos{(\omega t - k z)} ~\hat{x} \label{E_eq_plane_wave} 
\end{equation}
\begin{equation}
B ~=~ b_y \cos{(\omega t - k z)} ~\hat{y} \label{B_eq_plane_wave}. 
\end{equation}

The terms in the modified Maxwell's 
equations [equations~(\ref{Faraday_Eq}),~(\ref{Ampere_Eq})] for the above 
wave equations are computed below. 
\begin{equation}
\nabla \times B 
~ = ~ -b_yk\sin{(\omega{t}-k z)} ~~\hat{x}  \label{light_nabla_B}
\end{equation}
\begin{equation}
\nabla \times E 
~ = ~ e_xk\sin{(\omega{t}-k z)} ~~\hat{y} \label{light_nabla_E} 
\end{equation}
\begin{equation}
\frac{dB}{dt}~ =~  
\left[{- \omega  
~+kv_z }\right]b_y\sin{(\omega t - k z)}~\hat{y}
 \label{light_dt_B} 
\end{equation}
\begin{equation}
\frac{dE}{dt} ~ =~  
\left[{- \omega  
~+kv_z }\right]e_x\sin{(\omega t - k z)}
 ~\hat{x} \label{VdotnablaE}
\end{equation}

\noindent
Substituting equations~(\ref{light_nabla_B}-\ref{VdotnablaE}) into 
the modified Maxwell's equations [equations~(\ref{Faraday_Eq}), (\ref{Ampere_Eq})] 
yields 
\begin{equation}
e_x k   ~ = ~  -  \left[{
- \omega   
~+  k v_z  
}\right]b_y
 \label{Faraday_Eq_4} 
\end{equation}
\begin{equation}
-b_y k   ~ = ~   \epsilon  \mu 
\left[{ 
 -  \omega ~
+  k v_z    
}\right]e_x
  \label{Ampere_Eq_4}
\end{equation}
By setting $ e_x = b_yc $ 
and $ \mu \epsilon = \frac{1}{c^2}$ both the equations 
reduce to the common 
expression  
\begin{equation}
c k   ~ = ~   \left({
\omega   
~-  k v_z  
}\right) .  \label{light_speed_1}
\end{equation}
Note that the wave propagation speed $V_{light}$ is given by 
$\omega/k$; therefore,  the light propagation speed (in the 
z-direction) is additive, i.e., 
\begin{equation}
V_{light}    ~ = ~      \omega/k   ~~=~
c ~+ v_z  . 
\label{light_speed}
\end{equation}
Thus, the modified Maxwell's equations allow the nominal 
velocity of the field $V$, in which light is generated, 
to be added to the standard velocity of light when 
the field is  non-moving ---  
this follows directly from the invariance of modified 
Maxwell's equations. 
%In this sense, 
%light can be considered as a disturbance of a field, which 
%moves with velocity $c$ with respect to the nominal field.  

\vspace{0.1in}
It is noted that the Michelson-Morley experiment 
is expected to yield the null result with the moving fields approach
because  the velocity of light is constant in all directions  
with respect to frame of measurements 
(in which light was generated).

\subsection{Effect of star's velocity on aberration}
\begin{figure}[!ht]
\begin{center}
\includegraphics*[width=1in]{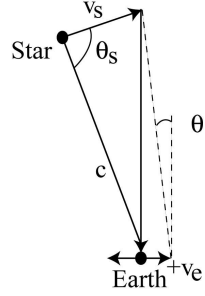}
\end{center}
\caption{Aberration formula based on relative-velocity matches the classical expression~\cite{Bradley1727}.}
\label{stellar_aberration}
\end{figure}

In a reference frame on earth, the velocity of the earth 
${V}_e = v_e \hat{V}_e$ adds to the velocity of stellar light 
to generate the aberration effect, see equation~(\ref{light_speed}),  
as in the original explanation by Bradley.~\cite{Bradley1727} \ 
The angle of the light direction with respect to earth  
($\theta$ measured perpendicular to earth's motion as 
shown in figure~\ref{stellar_aberration}) is maximum if 
the star's velocity ${V}_s= v_s \hat{V}_s$ reduces the nominal 
light speed to $c - v_s$ (when angle $\theta_s = 0$). 
Thus, the maximum change in the light direction with 
respect to earth is  
$2\theta$ where
\begin{equation}
\tan{\theta} ~ =~  \frac{v_e}{c - v_s}. 
\end{equation}

\noindent
For small speeds $v_s$ and $v_e$ the 
above expression is only linear in $v_e$ (and not linear in $v_s$) 
--- it can be approximated as 
\begin{equation}
\theta ~ \approx ~  \frac{v_e}{c} 
\end{equation}

The effect of the star's velocity ${V}_s$ on the 
aberration effect (due to Earth's motion) is small if the speed of the star is 
small, i.e., ${v}_s$ is much smaller than the nominal velocity of light $c$.  
Therefore, stellar aberration appears 
to be independent of the 
star's velocity $V_s$~\cite{Brecher_77} 
and appears to only depend on the relative change in the 
observer's velocity.~\cite{Phipps_89} \   
%a test to differentiate this.  

\subsection{Transverse Doppler effect}
\begin{figure}
\begin{center}
\includegraphics*[width=1.25in]{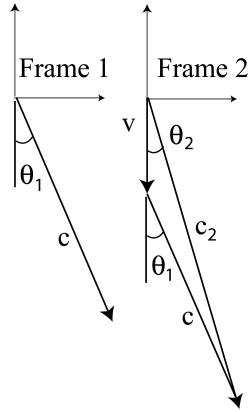}
\end{center}
\caption{Relative-velocity approach explains the transverse Doppler effect: Frame~1, which has relative velocity 
$V$ with respect to frame~2, generates light whose velocity 
is $c$ in frame~1. The observed velocity of the light 
is at an angle $\theta_2$ in frame~2} 
\label{Doppler_effect}
\end{figure}

% 
% Formula developed on the trip back from ACC in New York, 2007
%
Consider the Doppler effect due to addition of velocities in 
different frames. Let light  be generated in frame~1 with velocity 
$C = c \hat{C}$ and angle $\theta_1$ (in frame~1). Moreover, in frame~2, 
the observed velocity of light is $C_2 = {c}_2\hat{C}_2$ with angle $\theta_2$ 
as shown in Fig~\ref{Doppler_effect}. Furthermore, let frame~1 be moving 
with relative velocity ${V} = v \hat{V}$ relative to frame~2. 

\noindent
The magnitude ${c}_2$ of the observed velocity  
(at an angle $\theta_2$) can be determined 
from the law of cosines  
\begin{equation}
{v}^2 ~+ c_2^2 ~- 2{v}c_2 \cos{\theta_2}  =   c^2 
\end{equation}
which implies that $$ c_2 ~=  {v}cos{\theta_2} ~+\sqrt{c^2 
~-{v}^2\sin^2{\theta_2}}.$$ 

\noindent
Hence, the frequencies $f_1, f_2$ in the two frames are related by 
\begin{equation}
\begin{array}{rcl}
f_2 & = & f_1 \frac{c_2}{c} \\
& = &  f_1 \frac{v\cos{\theta_2} +\sqrt{{c}^2 
- {v}^2 \sin^2{\theta_2}}}{c} \\
& = & 
\left\{
\begin{array}{ll}
f_1 (1 +\frac{v}{c})  & ~{\mbox{if~}} \theta_2 = 0  \\
f_1 (1 -\frac{v}{c})  & ~{\mbox{if~}} \theta_2 = \pi  \\
f_1 \sqrt{1 -(\frac{v}{c})^2}  & ~{\mbox{if~}} 
\theta_2 = \pi/2   \label{doppler_shifts}.
\end{array}
\right.
\end{array}
\end{equation}
The expression for transverse Doppler effect, 
in equation~(\ref{doppler_shifts}) when $\theta_2 = \pi/2$,    
matches the relativistic expression (e.g., Born,~\cite{Born_62} \ page~301).

\subsection{Convection of light in moving media}
The effect of moving media on the velocity of light through the 
media is shown to be 
similar to Fresnel's drag formula without 
the need for Lorentz contraction that was developed 
to explain this effect. 

\begin{figure}[!ht]
\begin{center}
\includegraphics*[width=3in]{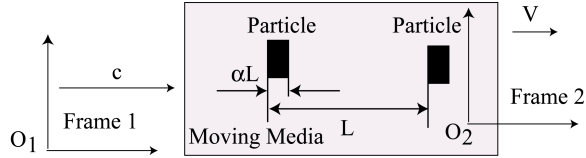}
\end{center}
\caption{Relative-velocity approach to model the convection effect (Fresnel drag) in moving media.  
In the moving media (frame~2) the  mean length  between 
particles is $L$ and the average length of each particle is $\alpha L$. 
%The speed of light through the particle is $c_m$ and 
%through the vacuum between the particles is $c$.}
}
\label{Convection_effect}
\end{figure}

\vspace{0.1in}
Consider a media moving with 
relative velocity $V = v\hat{V}$ 
in frame~1 as shown in figure~\ref{Convection_effect}. 
For an observer $O_1$ in frame~1, the 
speed of light generated in frame~1 is $c$ (in vacuum); 
the goal is to estimate the effective speed of light $c_{eff,O_1}$ 
through the moving media for the same observer (in frame~1). 
The passage of light in the moving media can be differentiated 
into two types: (a)~the passage of light through 
particles 
in the media;  and (b)~passage through vacuum in the 
media --- this approach is adapted from the method by Michelson and Morley.~\cite{Michelson_86} \  
Let the  mean length  between 
particles be $L$ and the mean length of each particle be 
$\alpha L$ --- these are measured in frame 2 that is fixed 
on the moving media (as shown in figure~\ref{Convection_effect}). 
It is noted that the positive factor $\alpha$ tends to be small, i.e., 
the particle length is small when compared to the distance 
between particles.~\cite{Michelson_86} \

Consider an observer $O_2$ in frame 2; let the nominal speed 
of light through a particle in the medium be $c_m$ when the 
relative velocity $V$ is zero. However, due to motion of the 
media, the speed of light (generated in frame 1) through  
particle is $c_m -v$ and through vacuum is $c-v$ for 
observer 2. The total velocity is not a linear summation of the two velocities; the effective speed of light 
$c_{eff, O_2}$ through the medium (for a fixed 
observer in frame $O_2$) is given by 
\begin{equation}
%\begin{array}{rcl}
\frac{L}{c_{eff, O_2}}~~=~  \frac{\alpha L}{c_{m}-v} 
~~+ \frac{(1-\alpha)L}{c-v}~~~
{\mbox{i.e.,~~~~}} {c_{eff, O_2}} ~= \frac{1}{ 
\frac{\alpha}{c_{m}-v} ~~+ \frac{(1-\alpha)}{c-v}}.
\label{moving_media_eff_vel}
%\end{array}
\end{equation}

\noindent
The nominal speed $C_{nom}$ of the light  through the 
media with zero relative velocity  is obtained by 
setting $v=0$ in equation~(\ref{moving_media_eff_vel}) as  
\begin{equation}
{c_{nom}} =  \frac{1}{ \frac{\alpha}{c_{m}} 
~~+ \frac{(1-\alpha)}{c}}.
\label{moving_media_nom_vel}
\end{equation} 

\noindent
The effective velocity expression in 
equation~(\ref{moving_media_eff_vel}) can be expanded in 
terms of the 
relative velocity $V$ as (where the higher order 
terms are neglected) 
\begin{equation}
\begin{array}{rcl}
{c_{eff, O_2}} & \approx & {c_{nom}} ~ + 
\frac{-1}{\left(\frac{\alpha}{c_{m}} ~~+ \frac{(1-\alpha)}{c}\right)^2} 
\left({ \frac{\alpha}{c_{m}^2} ~~+\frac{(1-\alpha)}{c^2}} 
\right) v   \\
& = & {c_{nom}} ~ - 
\frac{c_{nom}^2}{c^2}
\left({ \frac{\alpha c^2}{c_{m}^2} ~~+{(1-\alpha)}} \right) v  
 \\
& = & {c_{nom}} ~ - 
\frac{1}{\eta^2}
\left({ \frac{\alpha c^2}{c_{m}^2} ~~+{(1-\alpha)}} \right) v 
\label{moving_media_approx_vel}
\end{array}
\end{equation}
where $\eta$ is the media's coefficient of refraction. 
If $\alpha$ is small, then the 
expression in equation~(\ref{moving_media_approx_vel}) can be  
approximated by  
\begin{equation}
{c_{eff, O_2}} ~ \approx ~~ {c_{nom}} ~ - 
\frac{1}{\eta^2} v . 
\label{moving_media_approx_2}
\end{equation}

\noindent
Rewriting in terms of observer $O_1$ in frame~1, by adding $v$ 
to the expression, leads to 
\begin{equation} 
\begin{array}{rcl}
c_{eff, O_1} & =  & {c_{eff, O_2}} + v ~=~ {c_{nom}} ~ - 
\frac{1}{\eta^2} v  + v   \\ 
& = & 
{c_{nom}} + \left( 1 -  
\frac{1}{\eta^2} \right) v, 
\label{moving_media_Fresnel} 
\end{array}
\end{equation}
which is the same as Fresnel's expression. 

\section{Conclusions}
This article presented a Weber-type,   
relative-velocity-dependent electromagnetism model.   
It is shown that the model: 
(i)~captures relativistic effects in optics, high-energy particles, and gravitation; and  
(ii)~explains the apparent discrepancies in experimental energy measurements.


\begin{thebibliography}{99}


\bibitem{Weber_assis}
A.~K.~T. Assis and H.~T. Silva.
\newblock Comparison between weber's electrodynamics and classical
  electrodynamics.
\newblock {\em Pramana --- Journal of Physics}, 55(3):393--404, September 2000.

\bibitem{Turin_Crane_II}
J.~J. Turin and H.~R. Crane.
\newblock The absorption of high energy electrons, {Part~II}.
\newblock {\em Physical Review}, 52:610--613, September 15 1937.

\bibitem{Ellis_Wooster_1927}
C.~D. Ellis and W.~A. Wooster.
\newblock The average energy of disintegration of {Radium~E}.
\newblock {\em Proceedings of the Royal Society of London. Series A, Containing
  Papers of a Mathematical and Physical Character}, 117(776):109--123, December
  1 1927.

\bibitem{Madgwick}
E.~Madgwick.
\newblock The {$\beta$}-ray spectrum of {Ra E}.
\newblock {\em Proceedings of the Cambridge Philosophical Society},
  23:982--984, October 24 1927.

\bibitem{scott_35}
F.~A. Scott.
\newblock Energy spectrum of the beta-rays of {Radium E}.
\newblock {\em Physical Review}, 48:391--395, September 1 1935.

\bibitem{Ho_36}
P.~C. Ho and M.~H. Wang.
\newblock Beta-ray spectrum of {Radium E}.
\newblock {\em Chinese Journal of Physics}, 2(1):1--9, April 1936.

\bibitem{Slawsky_Crane_IV}
M.~M. Slawsky and H.~R. Crane.
\newblock The absorption of high energy electrons, {Part~IV}.
\newblock {\em Physical Review}, 59:1203--1210, December 15 1939.

\bibitem{Richardson36}
J.~R. Richardson and F.~N.~D. Kurie.
\newblock The radiations emitted from artificially produced radioactive
  substances.
\newblock {\em Physical Review}, 50:999--1006, 1, December 1936.

\bibitem{Martin_Townsend39}
L.~H. Martin and A.~A Townsend.
\newblock The {$\beta$}-ray spectrum of {Ra E}.
\newblock {\em Proceedings of the Royal Society of London. Series A, Containing
  Papers of a Mathematical and Physical Character}, 170(941):190--205, 1, March
  1939.

\bibitem{neary_40}
G.~J. Neary.
\newblock The {{$\beta$}-Ray Spectrum of Radium~E}.
\newblock {\em Proceedings of the Royal Society of London. Series A,
  Mathematical and Physical Sciences}, 175(960):71--87, March 28 1940.

\bibitem{Goldstein_80}
H.~Goldstein.
\newblock {\em Classical Mechanics}.
\newblock Addison-Wesley, Menlo Park, CA, second edition, 1980.

\bibitem{Thomson1897}
J.~J. Thomson.
\newblock Cathode rays.
\newblock {\em The London, Edinburgh, and Dublin Philosophical Magazine and
  Journal of Science, Fifth Series}, 44(269), October 1897 ({\em{reprinted in
  Classical Scientific Papers, Physics, by American Elsevier Publishing
  Company, Inc., New York, 1964, pp. 77-100}}).

\bibitem{Bradley1727}
James Bradley.
\newblock {A {L}etter from the {{R}}everend {M}r. {J}ames {B}radley {S}avilian
  {P}rofessor of {A}stronomy at {O}xford, and {F.R.S.} to {D}r. {E}dmond
  {H}alley {A}stronom. {R}eg., \& c. {G}iving an {A}ccount of a {N}ew
  {D}iscovered {M}otion of the {F}ix'd. {S}tars}.
\newblock {\em Philosophical Transactions}, 35:637--661, 1727.

\bibitem{Brecher_77}
Kenneth Brecher.
\newblock Is the speed of light independent of the velocity of the source?
\newblock {\em Physical Review Letters}, 39(17):1051--1054, 24 October 1977.

\bibitem{Phipps_89}
Thomas~R. {Phipps,~Jr.}
\newblock Relativity and aberration.
\newblock {\em American Journal of Physics}, 57(6):549--551, June 1989.

\bibitem{Born_62}
Max Born.
\newblock {\em Einstein's Theory of Relativity}.
\newblock Dover Publication, Inc., New York, revised edition, 1962.

\bibitem{Michelson_86}
A.~A. Michelson and E.~W. Morley.
\newblock Influence of motion of the medium on the velocity of light.
\newblock {\em American Journal of Science}, 31(185):377--386, May 1886.



\end{thebibliography}
\end{document}